\documentstyle[13lomcon,cite,epsfig]{article}

\def\MET{\mbox{${\hbox{$E$\kern-0.6em\lower-.1ex\hbox{/}}}_T$}} 
\def\ipb{pb$^{-1}$}                     

\def\MP{\mbox{$M_P$}}

\def\mbh{\mbox{$M_{\rm BH}$}\ }     
\def\MBH{\mbox{$M_{\rm BH}$}}         
\def\MET{\mbox{${\hbox{$E$\kern-0.6em\lower-.1ex\hbox{/}}}_T$}} 
\def\met{\mbox{${\hbox{$E$\kern-0.6em\lower-.1ex\hbox{/}}}_T$}\ } 
\def\ifb{fb$^{-1}$}                     

\def\METs{{\mbox{$E\kern-0.57em\raise0.19ex\hbox{/}_{T}$}}}
\def\METc{{\mbox{$E\kern-0.57em\raise0.19ex\hbox{/}_{T}^{\it cal}$}}\ }

\def\ipb{{pb$^{-1}$\ }}
\def\ifb{{fb$^{-1}$\ }}
\def\itev{{TeV$^{-1}$}}

\begin{document}

\title{COLLIDER SEARCHES FOR EXTRA SPATIAL DIMENSIONS AND BLACK HOLES}

\author{ Greg Landsberg \footnote{e-mail: landsberg@hep.brown.edu}}

\address{Brown University, Department of Physics, 182 Hope St., Providence, RI 02912, USA}


\maketitle\abstracts{
Searches for extra spatial dimensions remain among the most popular 
new directions in our quest for physics beyond the Standard Model.
High-energy collider experiments of the current decade should be able
to find an ultimate answer to the question of their existence in a variety 
of models. We review these models and recent
results from the Tevatron on searches for large, TeV$^{-1}$-size, and 
Randall-Sundrum extra spatial dimensions.
The most dramatic consequence of low-scale ($\sim 1$~TeV) quantum gravity is copious production of mini-black holes at the LHC. We discuss selected topics in the mini-black-hole phenomenology.}

\section{Models with Extra Spatial Dimensions}

A new, string theory inspired paradigm \cite{ADD} proposed by Arkani-Hamed, Dimopoulos, and Dvali (ADD) in 1998 suggested the solution to the hierarchy problem of the standard model (SM) by introducing several ($n$) spatial extra dimensions (ED) with the compactification radii as large as $\sim 1$~mm. These {\it large extra dimensions\/} are introduced to solve the hierarchy problem of the SM by lowering the Planck scale to a TeV energy range. (We further refer to this {\it fundamental\/} Planck scale in the (4+$n$)-dimensional space-time as $M_D$.) In this picture, gravity permeates the entire multidimensional space, while all the other fields are constrained to the 3D-space. Consequently, the {\it apparent\/} Planck scale $M_{\rm Pl} = 1/\sqrt{G_N}$ only reflects the strength of gravity from the point of view of a 3D-observer and therefore can be much higher than the fundamental (4+$n$)-dimensional Planck Scale. The size of large extra dimensions ($R$) is fixed by their number, $n$, and the fundamental Planck scale $M_D$. By applying Gauss's law, one finds~\cite{ADD,GRW}: $M^2_{\rm Pl} = 8\pi M_D^{n+2}\, R^n$. If one requires $M_D \sim 1$~TeV and a single extra dimension, its size has to be of the order of the radius of the solar system; however, already for two ED their size is only $\sim 1$~mm; for three ED it is $\sim 1$~nm, i.e., similar to the size of an atom; for larger number of ED it further decreases to subatomic sizes and reaches $\sim 1$~fm for seven ED.

Almost simultaneously with the ADD paradigm a very different low-energy utilization of the idea of compact extra dimensions has been introduced by Dienes, Dudas, and Gherghetta~\cite{DDG}. In their model, additional dimension(s) of the ``natural'' EWSB size of $R \sim 1 $~TeV$^{-1}$~\cite{itev} are added to the SM to allow for low-energy unification of gauge forces. In conventional SM and its popular extensions, such as supersymmetry, gauge couplings run logarithmically with energy, which is a direct consequence of the renormalization group evolution (RGE) equations. Given the values of the strong, EM, and weak couplings at low energies, all three couplings are expected to ``unify'' (i.e., reach the same strength) at the energy $\sim 10^{13}$~TeV, know as the Grand Unification Theory (GUT) scale. However, if one allows gauge bosons responsible for strong, EM, and weak interactions to propagate in extra dimension(s), the RGE equations would change. Namely, once the energy is sufficient to excite Kaluza-Klein (KK) modes of gauge bosons (i.e., $\sim 1/R \sim 1$~TeV), running of the couplings is proportional to a certain power of energy, rather than its logarithm. Thus, the unification of all three couplings can be achieved at much lower energies than the GUT scale, possibly as low as 10-100 TeV~\cite{DDG}. While this model does not incorporate gravity and thus does not explain its weakness relative to other forces, it nevertheless removes another hierarchy of a comparable size -- the hierarchy between the EWSB and GUT scales.

In 1999, Randall and Sundrum offered a rigorous solution~\cite{RS} to the hierarchy problem by adding a single extra dimension (with the size that can range anywhere from $\sim 1/M_{\rm Pl}$ virtually to infinity) with a non-Euclidean, warped metric. They used the Anti-deSitter (AdS) metric (i.e. that of a space with a constant negative curvature) $ds^2 = \exp(-2kR|\varphi|)\eta_{\mu\nu}dx^\mu dx^\nu - R^2d\varphi^2$, where $0 \le \varphi < 2\pi$ is the coordinate along the extra spatial dimension of radius $R$, $k$ is the curvature of the AdS space (warp factor), $x^\mu$ are the convential (3+1)-space-time coordinates, and $\eta^{\mu\nu}$ is the metric of the Minkowski space-time. A 3D-brane with positive tension is put at $\varphi = 0$. If gravity originates on this (Planck) brane, the wave function of the graviton has a peculiar feature that it is exponentially suppressed away from the brane in the direction of the extra dimension. If all the SM fields were confined to the Planck brane, one would not have seen any low-energy effects in this model. However, if a second (SM) brane with negative tension is put at $\varphi = \pi$, than the $M_{\rm Pl}$-size operators on the Planck brane would result in low-energy effects on the SM brane with the typical scale of $\Lambda_\pi = \overline{M}_{\rm Pl}\exp(-k\pi R)$, where $\overline{M}_{\rm Pl} \equiv M_{\rm Pl}/\sqrt{8\pi}$ is the reduced Planck mass. If the SM fields are confined to the SM brane, the hierarchy problem is solved for $\Lambda_\pi \sim 1$~TeV, which can be achieved with a little amount of fine tuning by requiring $kR \sim 10$. Since the only fundamental scale in this model is $M_{\rm Pl}$, the hierarchy problem is solved naturally for $R \sim 1/M_{\rm Pl}$. In the simplest Randall-Sundrum (RS) model~\cite{RS}, gravitons are the only particles propagating in the AdS space.

Numerous attempts to find large ED or constrain the ADD model have been carried out since 1998. They include measurements of gravity at short distances, studies of various astrophysical and cosmological implications of large ED, and numerous collider searches for virtual and real graviton effects. For detailed reviews of the existing constraints and the sensitivity of future experiments, see Ref. \cite{EDreviews}. The host of experimental measurements conducted to date have largely disfavored only the case of two or less large ED; for any larger number of them, the lower limit on the fundamental Planck scale is only $\sim 1$~TeV, hardly probing the most natural range of scales expected in the ADD model.

As was pointed out a decade ago \cite{BHearly}, an exciting consequence of TeV-scale quantum gravity is the possibility of producing black holes (BH) in high-energy interactions, accessible at colliders or by ultra-high-energy cosmic rays. More recently, this phenomenon has been quantified for the case of TeV-scale particle collisions\cite{dlgt}, resulting in a mesmerizing prediction that the LHC would produce mini black holes at an enormous rate (e.g., $\sim 1$~Hz for $M_D = 1$~TeV), thus becoming a black-hole factory. This observation led to an explosion of follow-up publications on the properties of mini-black holes produced in the lab and made this subject one of the most actively studied aspects of  phenomenology of models with extra dimensions. Here we review only some of the basic facts in phenomenology of black holes. For more extensive reviews, including the latest developments, see Ref. \cite{BHreviews}.

\section{Collider Searches for Extra Dimensions}

The most experimentally interesting feature of the above models with ED is rich low-energy phenomenology that originates from the KK spectrum of various particles propagating in ED. In what follows we compare the KK spectrum observed in the ADD, \itev, and RS models and discuss various experimental constraints on the model parameters.

In the ADD model, the only particle that propagates in ED and acquires KK modes is the graviton. Given the size of ED ($\sim 10^{-3}$--$10^{-15}$~m), energy spacing between the KK excitations of the graviton is $\sim 1$ meV -- 100~MeV. Consequently, the adjacent modes are hard to resolve and in most of the experiments the KK spectrum would appear continuous, from zero to a certain ultraviolet cutoff above which quantum gravity effects would modify this semi-classical picture. Since the fundamental Planck scale in the ADD model is $M_D \sim 1$~TeV, it's natural to expect this cutoff, $M_S$, to be of the same order: $M_S \sim M_D$. While each KK mode couples to the energy-momentum tensor with the gravitational strength $G_N$, the sheer number of the available KK modes is sufficient to enhance gravitational attraction tremendously.

In the \itev\ model with a single extra dimension of the size $R$, the zeroth KK mode of a gauge boson is the SM particle of mass $M_V$. This mass can be either $0$ (photon, gluon) or $>0$ ($W$, $Z$). The mass of the $i$-th KK mode is given by $M_i = \sqrt{M_V^2 + i^2/R^2}$. For the compactification scale $M_C \equiv 1/R \sim 1$~TeV, $1/R \gg M_V$ and hence the non-zeroth KK modes for all gauge bosons are nearly degenerate in mass, as the mass of the $i$-th mode is approximately equal to $i M_C$ for any $i > 0$. The KK tower of excited gauge bosons is expected to be truncated at the masses of the order of the new GUT scale ($\sim 100$~TeV), where new physics (e.g., brane dynamics) should take over.

Finally, in the Randall-Sundrum model the zeroth KK mode of the graviton remains massless, while the higher modes have masses spaced as subsequent zeros of the Bessel's function $J_1$. If the mass of the first excited mode of the graviton is $M_1$, the subsequent excitations would have masses of 1.83$M_1$, 2.66$M_1$, 3.48$M_1$, 4.30$M_1$, etc. The spacing between the adjacent KK excitations decreases at high masses. The zeroth excitation is coupled to the energy-momentum tensor with the gravitational strength $G_N$, while each of the higher excitations couples with the strength $\sim 1/\Lambda_\pi^2$, i.e., comparable to the EW coupling strength. The excited modes are presumably truncated at masses $\sim M_{\rm Pl}$, where brane excitations would modify the discussed behavior.

Since KK gravitons couple to the energy-momentum tensor, they can be added to any vertex or line of any SM Feynman diagram accessible at colliders. Consequently, to probe the ADD model one could look for direct emission of KK gravitons, e.g., in the $q \bar q \to g/\gamma + G_{\rm KK}$ process, which results in a single jet or photon ({\it monojet\/} or {\it monophoton\/}) in the observable final state. The experimental signature for such events is an apparent transverse momentum non-conservation and an overall enhancement of the tail of the jet/photon transverse energy spectrum. Another type of effect observable in high-energy collisions is an enhancement of Drell-Yan or diboson spectrum at high invariant masses due to additional diagrams involving virtual KK graviton exchange. Searches for virtual graviton effects are complementary to those for direct graviton emission, since the former depend on the ultraviolet cutoff of the KK spectrum, $M_S$, while the latter depends directly on the fundamental Planck scale $M_D$.

While direct graviton emission cross section is well defined~\cite{Peskin,GRW}, the cross section for virtual graviton exchange depends on a particular representation of the interaction Lagrangian and the definition of the ultraviolet cutoff $M_S$. We will consider two such representations~\cite{GRW,HLZ}. In both of them, the effects of ED are parameterized via a single variable $\eta_G = {\cal F}/M_S^4$, where ${\cal F}$ is a dimensionless parameter of order one reflecting the dependence of virtual $G_{\rm KK}$ exchange on the number of extra dimensions. Ref. \cite{GRW} simply fixes ${\cal F}$ at 1, while an attempt to resolve the $n$-dependence is made in~\cite{HLZ}. In both cases the interference with the SM diagrams is assumed to be constructive.

LEP experiments have pioneered searches for large extra dimensions at colliders in both direct graviton emission in the $e^+ e^- \to \gamma/Z + G_{\rm KK}$ channel and via virtual graviton effects in fermion pair, as well as in diboson production. The best sensitvity is achieved in the $e^+ e^- \to \gamma + G_{\rm KK}$ and in the $e^+ e^- \to e^+ e^-/\gamma\gamma$ channels. For a review of LEP limits see, e.g., Refs. \cite{EDreviews,LEPcomb}.

Both the CDF and D\O\ Collaborations at the Fermilab Tevatron $p\bar p$ collider sought for large ED since Run 1. D\O\ has pioneered searches for virtual graviton effects in the dielectron and diphoton channels at hadron colliders~\cite{D0-diem}, as well as searches in the challenging monojet channel~\cite{D0-monojet} plagued by copious instrumental backgrounds from jet mismeasurement and cosmics. CDF has pioneered searches in the monophoton channel at hadron colliders~\cite{CDF-monophoton} and also accomplished a monojet analysis~\cite{CDF-monojet}. 

Recently these results have been updated with 1--3 \ifb of data collected in Run 2 in the dimuon, dielectron, diphoton, monophoton, and monojet channels~\cite{ADD-Run2,monophotons-CDF,monophotons-D0,monojets-CDF,diem-D0} and also for the first time using the dijets~\cite{dijets}. These limits are listed in Tables~\ref{table:Run2} and \ref{table:virtual}. The tightest limits on $M_D$ come from the CDF combination of the monojet and monophoton channels and combined LEP limits. CDF limits exceed LEP limits for $n \ge 4$. The tightest limits on the ultraviolet cutoff $M_S$ come from the D\O\ experiment in the combined $ee+\gamma\gamma$ channel.

\begin{table}[t]
\caption{\label{table:Run2}
Most recent 95\% CL lower limits on the fundamental Planck scale $M_D$ (in TeV).}
\begin{center}
\begin{tabular}{lccccccc}
\hline
Experiment and channel 	& $n=2$ & $n=3$ & $n=4$ & $n=5$ & $n=6$ \\
\hline
LEP Combined~\protect\cite{LEPcomb} & 1.60 & 1.20 & 0.94 & 0.77 & 0.66 \\
\hline
CDF monophotons, 2.0 \ifb \protect\cite{monophotons-CDF}&   1.08    &  1.00   & 0.97 &  0.93  & 0.90  \\
D\O\ monophotons, 2.7 \ifb \protect\cite{monophotons-D0}&   0.97    &  0.90   & 0.87 &  0.85  & 0.83 \\
\hline
CDF monojets, 1.1 \ifb \protect\cite{monojets-CDF} & 1.31  & 1.08 & 0.98 & 0.91 & 0.88 \\
\hline
CDF combined \protect\cite{monophotons-CDF}& 1.42 & 1.16 & 1.06 & 0.99 & 0.95 \\
\hline
\end{tabular}
\end{center}
\vskip -0.2in
\end{table}

\begin{table}[t]
\caption{Recent 95\% CL lower limits on the ultraviolet cutoff $M_S$ (in TeV) from the Tevatron Run 2. NLO QCD effects have been accounted for via a SM $K$-factor.}
\label{table:virtual}
\begin{center}
\begin{tabular}{lc@{}cccccc@{}cc}
\hline
D\O\ Signature & GRW~\cite{GRW} & \multicolumn{6}{@{}c}{HLZ~\cite{HLZ}}\\
\hline
& & ~~$n$=2 & $n$=3 & $n$=4 & $n$=5 & $n$=6 & $n$=7 \\
\hline
$ee+\gamma\gamma$, 1.1 \ifb \protect\cite{diem-D0}& 1.62 & 2.09 & 1.94 & 1.62 & 1.46 & 1.36 & 1.29 \\
Dijets, 0.7 \ifb \protect\cite{dijets}                               & 1.56 &         & 1.85 & 1.56 & 1.41 & 1.31 & 1.24 \\
\hline
\end{tabular}
\end{center}
\vskip -0.1in
\end{table}

As of the beginning of Run 2, there have been no dedicated searches for \itev\ extra dimensions. However, a number of constraints have been derived by phenomenological analysis of the existing data. For a review of indirect cosntraints see, e.g., Ref.~\cite{KCGL2}. The best limits come from LEP electroweak precision measurements; the combined limit on the compactification scale of the \itev\ dimensions $M_C$ approaches 6.8 TeV for a single \itev\ extra dimension. By now, the \itev\ extra dimensions have been looked for via virtual gauge boson excitation effects in the dielectron (200 \ipb\cite{itev-D0}) and dijet (700 \ipb\cite{dijets}) channels at D\O. The preliminary lower 95\% CL limits on the compactification scale $M_C$ of a single extra dimension set in these analyses are 1.12 and 1.4 TeV respectively.

Both the CDF and D\O\ collaborations pioneered direct searches for the effects of Kaluza-Klein gravitons in the Randall-Sundrum model in Run 2~\cite{RS1,RS2}. As discussed above, the simplest RS model is fixed by specifying just two parameters: the curvature of the AdS space $k$ and the radius of compactification $R$. However, from the experimental point of view, it is more convenient to work with an equivalent set of parameters, which correspond to direct observables: $M_1$, the mass of the first Kaluza-Klein excitation of the graviton, and a dimensionless parameter $\tilde{k} = k/\overline{M}_{\rm Pl}$, which governs the coupling of the gravitons to the SM fields and hence defines the internal width of the KK gravitons. Both the CDF and D\O\ Collaborations used the dilepton and diphoton mass spectra and searched for a narrow resonance, which would indicate a production of the first excitation of the KK graviton. The best sensitivity comes from the 1~\ifb combined D\O\ $ee + \gamma\gamma$ analysis~\cite{RS1} and 2.5 \ifb CDF $\mu\mu$ analysis, which exclude RS gravitons with masses below 0.9, 0.7, and 0.3 TeV for $\tilde{k} = 0.1$, 0.05, and 0.01, respectively.

Ultimate sensitivity to large, \itev, and RS extra dimensions will be achieved very soon at the LHC. In most of the cases, just $\sim 1$ \ifb of data is sufficient to discover them or severely constrain parameters of these models. The LHC would also allow to search for the KK excitations of the gauge bosons directly. Undoubtedly, the most exciting possibility is production of mini-black holes at the LHC~\cite{dlgt} possible in the ADD and RS models.

\section{Black Holes at the LHC}

Consider two partons with the center-of-mass energy $\sqrt{\hat s} =
\MBH$ colliding head-on. If the impact parameter of the collision is less than the (higher dimensional) Schwarzschild radius, corresponding to this energy, a black hole (BH) with the mass \mbh is formed. Therefore the total cross section of black hole production in particle collisions can be estimated from pure geometrical arguments and is of order $\pi R_S^2$. BH production is expected to be a threshold phenomena and the onset is expected to happen for a minimum black hole mass $\sim  M_D$ (ADD) or $\Lambda_\pi$ (RS). The total production cross section above this threshold at the LHC, ranges between 15 nb and 1 pb for the Planck scale between 1 TeV and 5 TeV, and varies by less than a factor of two for $n$ between 1 (RS) and 7. Note that this cross section is comparable with, e.g., $t\bar t$ production cross section, which result in $\sim 1$~Hz signal event rate at the nominal LHC luminosity. 

Once produced, mini black holes quickly ($\sim 10^{-26}$~s) evaporate via Hawking 
radiation~\cite{Hawking} with a characteristic temperature
of $\sim 100$~GeV~\cite{dlgt}. The average multiplicity of particles 
produced in the process of BH evaporation is given by\cite{dlgt} and 
is of the order of half-a-dozen for typical BH masses accessible 
at the LHC. Since gravitational coupling is flavor-blind, a BH emits all the 
$\approx 120$ SM particle and antiparticle degrees of freedom with 
roughly equal probability. Accounting for color and spin, we expect 
$\approx 75\%$ of particles produced in BH decays to be quarks and gluons, 
$\approx 10\%$ charged leptons, $\approx 5\%$ neutrinos, and $\approx 5\%$ 
photons or $W/Z$ bosons, each carrying hundreds of GeV of energy. 

A relatively large fraction of prompt and energetic photons, electrons, 
and muons expected in the high-multiplicity BH decays would 
make it possible to select pure samples of BH events, which are also 
easy to trigger on~\cite{dlgt}. The reach of a simple 
counting experiment extends up to $\MP \approx 9$~TeV ($n=2$--7),
for which one would expect to see a handful of BH events with negligible 
background. 

\section{Conclusions}

If TeV-scale gravity is realized in nature, its detection via Kaluza-Klein effects as well as the possibility of production and detailed studies of black holes in the lab are just a few years away. That would mark an exciting transition for astroparticle physics: its true unification with cosmology~--- the ``Grand Unification'' to strive for.

\section*{Acknowledgments}

I would like to thank the organizers of the 13th Lomonosov Conference on Elementary Particle Physics for a kind invitation, warm hospitality, and a great conference. This work has been partially supported by the U.S.~Department of Energy under Grant No. DE-FG02-91ER40688 and the National Science Foundation under the CAREER Award PHY-0239367. These proceedings were written while I was visiting Aspen Center for Physics, which I would like to thank for its hospitality and support.


\section*{References}
\begin{thebibliography}{20}

\bibitem{ADD}
N.Arkani-Hamed, S.Dimopoulos, and G.Dvali, Phys. Lett. B {\bf 429}, 263 (1998); 
Phys. Rev. D {\bf 59}, 086004 (1999);
I.Antoniadis, N.Arkani-Hamed, S.Dimopoulos, and G.Dvali, Phys. Lett. B {\bf 436}, 257 (1998); 
N.Arkani-Hamed, S.Dimopoulos, and J.March-Russell, hep-th/9809124 (1998).

\bibitem{GRW}
G.Giudice, R.Rattazzi, and J.Wells, Nucl. Phys. {\bf B544}, 3 (1999), and a revised version hep-ph/9811291v2 (2000).

\bibitem{DDG}
K.R.Dienes, E.Dudas, and T.Gherghetta, Phys. Lett. B {\bf 436}, 55 (1998); 
Nucl. Phys. B {\bf 537}, 47 (1999); {\it ibid.\/} {\bf 567}, 111 (2000).

\bibitem{itev}
I.Antoniadis, K.Benakli, and M.Quiros, Phys. Lett. B {\bf 460}, 176 (1999).

\bibitem{RS}
L. Randall and R. Sundrum, Phys. Rev. Lett. {\bf 83}, 3370 (1999);
{\it ibid.\/} {\bf 83}, 4690 (1999).

\bibitem{EDreviews}
D.Bourilkov, hep-ex/0103039 (2001); 
J.L.Hewett and M.Spiropulu, Ann. Rev. Nucl. Part. Sci. {\bf 52}, 397 (2002);
G.Landsberg, hep-ex/0412028 (2004).

\bibitem{BHearly}
P.C.Argyres, S.Dimopoulos, and J.March-Russell, Phys. Lett. {\bf B441}, 96 (1998);
T.Banks and W.Fischler, JHEP {\bf 9906}, 014 (1999);
R.Emparan, G.T.Horowitz, and R.C.Myers, Phys. Rev. Lett. {\bf 85}, 499 (2000).

\bibitem{dlgt}
S.Dimopoulos and G.Landsberg, Phys. Rev. Lett. {\bf 87}, 161602 (2001); 
S.B.Giddings and S.Thomas, Phys. Rev. D {\bf 65}, 056010 (2002).

\bibitem{BHreviews}
M.~Cavaglia, Int. J. Mod. Phys. A {\bf 18}, 1843 (2003);
G.~Landsberg, Eur. Phys. J. C {\bf 33}, S927 (2004);  J. Phys. G {\bf 32}, R337 (2006);
P.~Kanti, Int. J. Mod. Phys. A {\bf 19}, 4899 (2004); arXiv:0802.2218 (2008).

\bibitem{Peskin}
E.A.Mirabelli, M.Perelstein, and M.E.Peskin, Phys. Rev. Lett. {\bf 82}, 2236 (1999).

\bibitem{HLZ}
T.Han, J.Lykken, and R.Zhang, Phys. Rev. D {\bf 59}, 105006 (1999) and a revised version hep-ph/9811350v4 (2000).

\bibitem{LEPcomb}
S.Ask, hep-ex/0410004 (2004).

\bibitem{D0-diem}
B.Abbott {\it et al.\/} (D\O\ Collaboration), Phys. Rev. Lett. {\bf 86}, 1156 (2001).

\bibitem{D0-monojet}
V.M.Abazov {\it et al.\/} (D\O\ Collaboration), Phys. Rev. Lett. {\bf 90}, 251802 (2003).

\bibitem{CDF-monophoton}
T.Affolder {\it et al.\/} (CDF Collaboration), Phys. Rev. Lett. {\bf 89}, 281801 (2002).

\bibitem{CDF-monojet}
T.Affolder {\it et al.\/} (CDF Collaboration), Phys. Rev. Lett. {\bf 92}, 121802 (2004).

\bibitem{ADD-Run2}
V.M.Abazov {\it et al.\/} (D\O\ Collaboration), Phys. Rev. Lett. {\bf 95}, 161602 (2005); 
{\it ibid.\/} {\bf 101} 011601 (2008);
A.Abulencia {\it et al.\/} (CDF Collaboration), Phys. Rev. Lett. {\bf 97}, 171802 (2006).

\bibitem{monophotons-CDF}
CDF Collaboration, {\tt http://www-cdf.fnal.gov/physics/exotic/r2a /20071213.gammamet/LonelyPhotons/photonmet.html} (2007).

\bibitem{monophotons-D0}
D\O\ Collaboration, D\O\ Note 5729-CONF, {\tt http://www-d0.fnal.gov/ Run2Physics/WWW/results/prelim/NP/N63/N63.pdf} (2008).

\bibitem{monojets-CDF}
CDF Collaboration, {\tt http://www-cdf.fnal.gov/physics/exotic/r2a /20070322.monojet/public/ykk.html} (2007).

\bibitem{diem-D0}
D\O\ Collaboration, to be submitted to Phys. Rev. Lett.;
O.Stelzer-Chilton, talk at the 34th Int. Conf. on High Energy Physics, ICHEP 08, {\tt http://www.hep.upenn.edu/ichep08/talks/misc/schedule} (2008).

\bibitem{dijets}
D\O\ Collaboration, D\O\ Note 5729-CONF, {\tt http://www-d0.fnal.gov/ Run2Physics/WWW/results/prelim/QCD/Q11/Q11.pdf} (2008).

\bibitem{KCGL2}
K.Cheung and G.Landsberg, Phys. Rev. D {\bf 65}, 076003 (2002).

\bibitem{itev-D0}
D\O\ Collaboration, D\O\ Note 4349-CONF, {\tt http://www-d0.fnal.gov/ Run2Physics/WWW/results/prelim/NP/N02/N02.pdf} (2004).

\bibitem{RS1}
V.M.Abazov {\it et al.\/} (D\O\ Collaboration), Phys. Rev. Lett. {\bf 95}, 091801 (2005); {\it ibid.\/}
{\bf 100}, 091802 (2008);

\bibitem{RS2}
A.Abulencia {\it et al.\/} (CDF Collaboration), Phys. Rev. Lett. {\bf 95}. 252001 (2005);
A.Aaltonen {\it et al.\/} (CDF Collaboration), Phys. Rev. Lett. {\bf 99}, 171801 (2007); 
{\it ibid.\/} {\bf 99}, 171802 (2007); arXiv:0801.1129 (2008);
CDF Note 9160-PUB {\tt http://www-cdf.fnal.gov/physics/exotic/ r2a/20080306.dielectron\_duke/pub25/cdfnote9160\_pub.pdf} (2008); CDF Collaboration, 
{\tt http://www-cdf.fnal.gov/physics/exotic/r2a/ 20080710.dimuon\_resonance/} (2008).

\bibitem{Hawking}
S.W.Hawking, Commun.  Math. Phys. {\bf 43}, 199 (1975).

\end{thebibliography}
\end{document}
